# The network paradigm as a modeling tool in regional economy: the case of interregional commuting in Greece

**Dimitrios Tsiotas[1]\*, Labros Sdrolias[2], Dimitrios Belias[2]**

[1]Department of Planning and Regional Development, University of Thessaly,
Pedion Areos, Volos, 38 334, Greece,
[2] Department of Business Administration, Technological Educational Institute of Thessaly,
Nea Ktiria, 411 10 Larissa, Greece,
Tel. +30 24210 74446, fax: +30 24210 74493
E-mails: tsiotas@uth.gr; lsdrolias@teilar.gr; dbelias@pe.uth.gr
*\* Corresponding author*

**Abstract**
Network Science is an emerging discipline using the network paradigm to model communication systems as pair-sets of interconnected nodes and their linkages (edges). This paper applies this paradigm to study an interacting system in regional economy consisting of daily road transportation flows for labor purposes, the so-called commuting phenomenon. In particular, the commuting system in Greece including 39 non-insular prefectures is modeled into a complex network and it is studied using measures and methods of complex network analysis and empirical techniques. The study aims to detect the structural characteristics of the Greek interregional commuting network (GCN) and to interpret how this network is related to the regional development. The analysis highlights the effect of the spatial constraints in the structure of the GCN, it provides insights about the major road transport projects constructed the last decade, and it outlines a population-controlled (gravity) pattern of commuting, illustrating that high-populated regions attract larger volumes of the commuting activity, which consequently affects their productivity. Overall, this paper highlights the effectiveness of complex network analysis in the modeling of systems of regional economy, such as the systems of spatial interaction and the transportation networks, and it promotes the use of the network paradigm to the regional research.

## 1. Introduction
Commuting is a multidimensional phenomenon concerning daily mobility for labor purposes outside the place of residence (Polyzos, 2011). The theoretical framework of this phenomenon has social, economic, geographical, and political dimensions, so that the study and the in-depth knowledge of commuting suggests a very complex procedure that can provide useful insights contributing to a more effective policy, especially in the fields of labor and transportation, but also in the sustainable transport planning (Evans et al., 2002; Van Ommeren and Rietveld, 2005). Up today, researchers have studied a wide range of commuting issues, such as transportation (spatial and time) costs (Van Ommeren and Fosgerau, 2009; Tsiotas and Polyzos, 2013a), psychology of mobility (Koslowsky et al., 1995), the possibility of a traffic accident (Ozbay et al., 2007), various issues about transportation mode and alternative routes selection (Murphy, 2009; Liu and Nie, 2011), and issues related to the relationship between commuting and forms of productivity (Van Ommeren and Rietveld, 2005).

However, the macroscopic study of commuting has not enjoyed so much attention, both in international and in national level, such as in the case of Greece (Tsiotas and Polyzos, 2013a; Polyzos et al., 2014). One of the modern scientific fields that is capable of providing modeling methods towards this holistic direction is the so-called complex network analysis (Brandes and Erlebach, 2005; Easley and Kleinberg, 2010; Barthelemy, 2011) or the recently renamed discipline of Network Science (Brandes et al., 2013). This approach models communication systems as graphs (Easley and Kleinberg, 2010; Borgatti and Halgin, 2011; Tsiotas and Polyzos, 2013a), namely as pair sets of interconnected units (nodes) and their connections (links or edges). According to this perspective, a commuting system can be modeled as a network, where (on an interregional level) nodes represent places of origin and destination whereas links include distance and flow information.

Within this conceptual framework, this paper models as a complex network the interregional commuting system in Greece, which is developed between capital cities of the Greek prefectures, and it studies the topology and functionality this network, in accordance to its socioeconomic environment. Also, an empirical analysis is applied and constructs a multivariate linear regression model to describe the interregional Greek commuting, based on the semantic components of the network concept, as described by Berners-Lee et al. (2007), Easley and Kleinberg (2010), and Tsiotas and Polyzos (2015c). The further purpose of the study is to detect the structural features of the commuting phenomenon, as these are reflected in the mobility captured by land transport infrastructures.

The remainder of this paper is organized as follows: Section 2 presents the methodological framework and particularly the modeling procedure and assumptions, the measures used in the network analysis and details of the empirical model constructed. Section 3 presents the results of the analysis and discusses them in the light of complex network analysis and regional economics, emphasizing on the transport sector. Finally, in Section 4 conclusions are given.

**2. Methodological framework**
*2.1. Network modeling*
The Greek Commuting Network (GCN) (Fig. 1) is a network with more an economic and less a physical interpretation. This spatial network represents an aspect of the national road network expressed on an interregional level, in which the routing and the roads geometry is not preserved, but only the geographical scale amongst city distances. Therefore, the GCN essentially represents the functionality of the road connectivity in Greece, in order to study the topology and the economic dynamics shaped by this system of spatial and economic interaction.

In particular, the GCN is represented in the *L*-space representation (Barthelemy, 2011; Tsiotas and Polyzos, 2015a,b), as a undirected spatial network $G(V,E)$, where the node-set ($V$) represents the capital cities of the Greek prefectures and the edge-set ($E$) expresses the potential of developing direct road connections between prefectures of Greece. The nodes' positions of the GCN on the map (Fig. 1) correspond to the geographical coordinates of the capital cities considered in the model, whereas the edge lengths are drawn proportionally (on a scale) and represent the road kilometric distances between pairs of nodes. Using capital cities to modeling the GCN is a decision of economic interpretation because capital cities of regions are spatial units with significant population concentration (Polyzos et al., 2013; Tsiotas and Polyzos, 2013a, b) and thus the GCN as a spatial network is a model with a significant economic (gravity) impact.

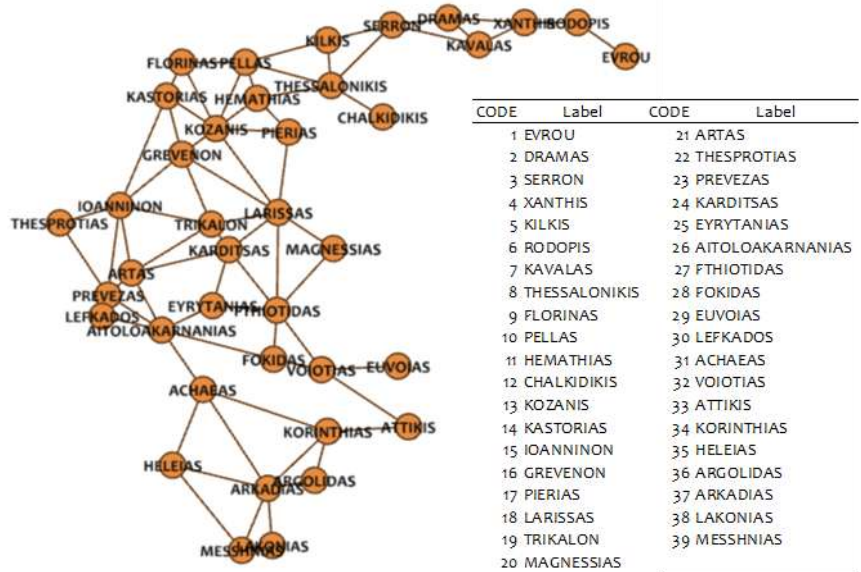

**Figure 1.** The Greek interregional commuting network (GCN), modeled in the *L*-space representation as a non-directed graph, with *n*=39 nodes and *m*=71 edges (nodes in the graph represent the capital cities of the prefectures defined by the Kapodestrian act).

The GCN is a connected network (including one component), consisting of *n*=39 mainland prefectures' capital cities (nodes) and *m*=71 interregional road links (edges) (Fig. 1). The edge spatial weights $ws_{,ij}=d(e_{ij})$ express the actual kilometric distances of the shortest paths connecting the capital cities of the county. Each edge represent two-way road sections and thus the total network is undirected with a symmetric adjacency matrix. Additional weights in GCN are time distances between the nodes, which express the required time (in minutes) to travel between two network nodes. These values suggest indirect indicators of the inter-regional road network effectiveness, since the average time of a route represents the quality of the network's road infrastructure (Tsiotas et al., 2012; Tsiotas and Polyzos, 2013a,b).

The spatial data (geographic coordinates) used for the construction of the GCN were extracted from Google Digital Mapping Services (Google maps, 2013), while the space-time data from the works of Tsiotas et al. (2012), Tsiotas and Polyzos (2013a,b), and Tsiotas and Polyzos (2015c). Available space-time data correspond to two time snapshots of the Greek interregional road network; the first concern the year 1988 and describe the past status of the national road transport, whereas the concern the year 2010 figures and represent a modern image of this network, resulted after the construction of some major national road infrastructure works, such as the Rio-Antirrio bridge and the Egnatia Motorway.

Finally, the GCN's commuting information was included into the spatial model in the form of node weights (Tsiotas and Polyzos, 2013a,b). That is, the available commuting data does not describe commuting flows on the network edges, but the commuting population mass originating from a certain county capital. This approach was considered as more representative in the regional level than using a model with an edge-weighted commuting flows, because the certain (node-weighted commuting) GCN model has available information for the total number of outgoing commuters capital city (even for those that have destinations other, non-capital, cities), whereas in the edge-weighted commuting model the commuting data will be restricted on just only between capital cities.

*2.2. Complex network measures*

Mixing space and topology measures are used in the GCN analysis and they are summarized in Table 1.

**Table 1**
Mixing space and topology measures used in the analysis of the GCN

| Measure | Description | Math Formula | Reference |
|---|---|---|---|
| Graph density ($\rho$) | Fraction of the existing connections of the graph to the number of the possible connections. It expresses the probability to meet in the GMN a connected pair of nodes. | $\rho = m \Big/ \binom{n}{2} = \dfrac{2m}{n \cdot (n-1)}$ | (Diestel, 2005) |
| Node Degree ($k$) | Number of the edges adjacent to a given node, expressing the node's communication potential. | $k_i = k(i) = \sum_{j \in V(G)} \delta_{ij}$, where $\delta_{ij} = \begin{cases} 1, \text{ if } e_{ij} \in E(G) \\ 0, \text{ otherwise} \end{cases}$ | (Diestel, 2005) |
| Node (spatial) strength ($s$) | The sum of edge distances being adjacent to a given node. | $s_i = s(i) = \sum_{j \in V(G)} \delta_{ij} \cdot d_{ij}$, where $d_{ij} = w(e_{ij})$ in km | (Barthelemy, 2011) |
| Average Network's Degree $\langle k \rangle$ | Mean value of the node degrees $k(i)$, with $i \in V(G)$. | $\langle k \rangle = \dfrac{1}{n} \cdot \sum_{i=1}^{n} k(i)$ | (Diestel, 2005) |
| Closeness Centrality ($CC(i)$) | Total binary distance $d(i,j)$ computed on the shortest paths originating from a given node $i \in V(G)$ with destinations all the other nodes $j \in V(G)$ in the network. This measure expresses the node's reachability in terms of steps of separation. | $CC(i) = \dfrac{1}{n-1} \cdot \sum_{j=1, i \neq j}^{n} d_{ij} = \overline{d}_i$ | (Koschutzki et al., 2005). |
| Betweenness Centrality ($CB(k)$) | The proportion of the ($\sigma$) shortest paths in the network that pass through a given node $k$. | $CB(k) = \sigma(k)/\sigma$ | (Koschutzki et al., 2005) |
| Local Clustering Coefficient ($C(i)$) | Probability of meeting linked neighbors around a node, which is equivalent to the number of the node's connected neighbors $E(i)$ (i.e. the number of triangles), divided by the number of the total triplets shaped by this node, which equals to $k_i(k_i-1)$. | $C(i) = \dfrac{E(i)}{k_i \cdot (k_i - 1)}$ | (Barthelemy, 2011) |
| Modularity ($Q$) | Objective function expressing the potential of a network to be subdivided into communities. In its mathematical formula, $g_i$ is the community of node $i \in V(G)$, $[A_{ij} - P_{ij}]$ is the difference of the actual minus the expected number of edges falling between a particular pair of vertices $i,j \in V(G)$, and $\delta(g_i,g_j)$ is an indicator function returning 1 when $g_i=g_j$. | $Q = \dfrac{\sum_{i,j}[A_{ij} - P_{ij}] \cdot \delta(g_i, g_j)}{2m}$ | (Blondel et al., 2008; Fortunato, 2010) |
| Average Path Length $\langle l \rangle$ | Average length $d(i,j)$ of the total of network shortest paths. | $\langle l \rangle = \dfrac{\sum_{v \in V} d(v_i, v_j)}{n \cdot (n-1)}$ | (Barthelemy, 2011) |

In addition to these fundamental network measures shown in Table 1, we calculate the omega (ω) index of Telesford et al. (2011), in order to empirically detect whether the GCN has small-world (SW) (Watts and Strogatz, 1998), lattice-like, or random-like characteristics. The SW property is rigorously defined on an available family of graphs, by detecting that the average path length scales logarithmically as the number of nodes tends to infinity, namely $\langle l \rangle = O(\log n)$, with $n \to \infty$ (Porter, 2012). Due to the unavailability of studying a family of graphs in most of the empirical cases, since it is quite data-demanding to collect many aspects of the same network for different time periods, the small-world attribute detection for the GCN is applied using the approximation $\omega$ index of Telesford et al. (2011). This index compares the clustering of the empirical network with that of a $p(k)$-equivalent (i.e. with the same degree distribution) lattice network $(\langle c \rangle_{latt})$ and the empirical network's path length with that of an $p(k)$-equivalent random network $(\langle l \rangle_{rand})$, according to the formula:

$$\omega = (\langle l \rangle_{rand} / \langle l \rangle) - (\langle c \rangle / \langle c \rangle_{latt}) \qquad (1)$$

The null models are computed using a randomization algorithm (Maslov and Sneppen, 2002) and the ''latticization'' algorithm (Rubinov and Sporns, 2010), which both preserve the degree distribution of the original network. Values of $\omega$ are restricted to the interval [-1,1], where those close to zero illustrate the SW attribute, positive values indicate *random characteristics*, whereas negative values indicate more *regular* or *lattice-like* characteristics (Tsiotas and Polyzos, 2015b).

*2.3. Empirical Analysis*

In this section we build an empirical model for describing the number of commuters in the GCN, by using network node-variables. Each variable is a collection of values corresponding to the $n=39$ in number network nodes, measured for a specific feature $p$. Thus each variable has 39 elements (equal to the number of network nodes) and refers to a single attribute (e.g. degree or population) (Tsiotas and Polyzos, 2013a, 2015c). Given that nodes in the GCN correspond to the Greek prefectures, $p=30$ in number vector-variables ($Y$, $X_1$, ..., $X_{29}$) are created and entered in the empirical analysis. The selection of the variables is made based on their relevance with the commuting phenomenon in the literature (Glaeser and Kohlhase, 2003, Clark et al., 2003, Ozbay et al., 2007, Van Ommeren and Fosgerau, 2009, Murphy, 2009; Liu and Nie, 2011; Polyzos, 2011; Tsiotas and Polyzos, 2015c), depending on the availability of the data. Furthermore, the variables used in the model are grouped into three thematic categories, a structural (S), a functional or behavioral (B), and an ontological (O), in order to comply with all three components that conceptually describe a network, as proposed by Tsiotas and Polyzos (2015c).

Within this context, all the 30 variables participating in the empirical analysis of GCN are shown in Table 2. The empirical model is constructed to describe the number of outgoing commuters per capital city as a function of the remaining 29 variables. Pearson's bivariate coefficient of correlation and a multivariate linear regression analysis (Norusis, 2004; Devore and Berk, 2012; Tsiotas and Polyzos, 2015c) are used to construct the model.

**Table 2**
Vector variables participating in the empirical analysis of the GCN

| Variable[*] | Description | Reference/Source |
|---|---|---|
| (*Structural Class* (**X**$_S$)) | | |
| ($S_1$) Road network degree | The number of connections being adjacent to each network node in the GCN. | (Tsiotas and Polyzos, 2015c; Google Maps, 2013) |

| Variable[*] | Description | Reference/Source |
|---|---|---|
| ($S_2$) Commuting degree | The number of commuting destinations per node in the GCN. | (Tsiotas and Polyzos, 2015c; ESYE, 2007) |
| ($S_3$) Degree difference | The difference $S_3 = S_1 - S_2$ | (Tsiotas and Polyzos, 2015c) |
| ($S_4$) Closeness centrality | Accessibility variable defined by the measure of closeness centrality shown in Table 1. | (Tsiotas et al., 2013c) |
| ($S_5$) Mobility centrality | Centrality measure that is analogue to the kinetic energy formula and captures the potential that a node attribute induces to the network. | (Tsiotas and Polyzos, 2013a, 2015a) |
| ($S_6$) Population | Population of each prefecture for the year 2011. | (Tsiotas and Polyzos, 2015a,b,c) |
| ($S_7$) Commuting sign | The sign computed from the statistical difference outgoing-incoming commuters per capital city (+: expelling, 0: neutral, -: attractive) | (Tsiotas and Polyzos, 2015c) |
| ($S_8$) Minimum commuting distance | Distance of each city's closest commuting destination. | (Tsiotas and Polyzos, 2015c) |
| ($S_9$) Average Neighbour Distance | Average distance of the $S_2$ variable's destinations. | (Tsiotas and Polyzos, 2015c) |
| ($S_{10}$) Bus Route Destinations | Number of available bus trip destinations per capital city. | (Tsiotas and Polyzos, 2015c) |
| ($S_{11}$) Train Route Destinations | Number of available train trip destinations per capital city. | (Tsiotas and Polyzos, 2015c) |
| (*Functional Class* ($\mathbf{X}_B$) | | |
| ($Y$)[**] Number of commuters | Number of commuters originating from each capital city. | (ELSTAT, 2011) |
| ($B_1$) Direct Commuters | Composite variable defined as $B_1=\max\{\text{incoming, outgoing commuters}\} \cdot S_7$. | (Tsiotas and Polyzos, 2015c) |
| ($B_2$) Average Bus Route Frequency | Each city's average number of weekly bus routes for all available destinations. | (Tsiotas and Polyzos, 2015c) |
| ($B_3$) Bus Flow Index | Is the product: $B_3 = B_2 \cdot S_{10}$ | (Tsiotas and Polyzos, 2015c) |
| ($B_4$) Average Train Route Frequency | Each city's average number of weekly train routes for all available destinations. | (Tsiotas and Polyzos, 2015c) |
| ($B_5$) Train Flow Index | Is the product: $B_5 = B_4 \cdot S_{11}$. | (Tsiotas and Polyzos, 2015c) |
| ($B_6$) Car Number | Number of privet cars at each prefecture. | (Tsiotas and Polyzos, 2015c) |
| ($B_7$) Bus Number | Number of buses at each prefecture. | (Tsiotas and Polyzos, 2015c) |
| ($B_8$) Taxi Number | Number of taxies at each prefecture. | (Tsiotas and Polyzos, 2015c) |
| *Ontological Class* ($\mathbf{X}_O$) | | |
| ($O_1$) Labor Population Percentage | Population percentage of each prefecture including ages 20<A<65 (extracted from 2011 consensus). | (Tsiotas and Polyzos, 2015c) |
| ($O_2$) Educational Index | Composite index of each prefecture's educational level. | (Polyzos, 2011) |
| ($O_3$) GDP | Gross Domestic Product of each prefecture. | (Polyzos, 2011; Tsiotas and Polyzos, 2015c) |
| ($O_4$) Welfare Index | Composite index showing each prefecture's welfare lever. | (Tsiotas and Polyzos, 2015c) |
| ($O_5$) Public Servants | The number of public servants of each | (Tsiotas and Polyzos, |

| Variable* | Description | Reference/Source |
|---|---|---|
| | prefecture. | 2015c) |
| ($O_6$) Transportations GDP | Each prefectures's GDP in the transportation sector. | (Tsiotas and Polyzos, 2015c) |
| ($O_7$) Accidents | Number of accidents per prefecture. | (Tsiotas and Polyzos, 2015c) |
| ($O_8$) Accidents Percentage | Per capita number of accidents for each perfecture ($O_8=O_7/S_6$). | (Tsiotas and Polyzos, 2015c) |
| ($O_9$) Productivity Dynamism | Composite index depending from the GDP change, the unemployment percentage, the labor productivity, and the labor percentage. | (Polyzos, 2011) |
| ($O_{10}$) Unemployment Inequalities | Inequalities index for unemployment per prefecture. | (Tsiotas and Polyzos, 2013a) |

\*. Symbols of the variables are given within parentheses.
\*\*. The *Y*-symbol is used because it participates as the response variable in the model

The algorithm of the model's construction consists of three steps. The first involves the grouping of the available variables into the three categories (structural, behavioral or functional and ontological) shown in Table 2. This process leads to the configuration of three sets of variables ($\mathbf{X}_S$, $\mathbf{X}_B$ και $\mathbf{X}_O$), according to the expression (Tsiotas and Polyzos, 2015b):

$$\mathbf{X} \equiv \{X_k, k=1,...,p\}$$
$$(\mathbf{X} \equiv \mathbf{X}_S \cup \mathbf{X}_B \cup \mathbf{X}_O) \wedge (\mathbf{X}_i \cap \mathbf{X}_j = \varnothing) \wedge (j \neq i, j = \{S,B,O\}) \quad (2)$$

where the sets $\mathbf{X}_S$, $\mathbf{X}_B$, and $\mathbf{X}_O$ represent the structural, functional, and ontological group, respectively.

In the second step, the algorithm distinguishes the most representative variables from each category, using the bivariate Pearson's coefficient of correlation $r(x,y) \equiv r_{xy}$. The representative variables for each category are those with the largest sum of squares of their correlation coefficients amongst all other variables (only statistically significant correlations are taken into account, $\alpha \leq 10\%$), which are computed (a) within a group $\mathbf{X}_{k=\{S,B,O\}}$ (within-groups calculations and (b) along all the groups for the total of $p=30$ available variables (global calculations). This process is expressed as follows (Tsiotas and Polyzos, 2015c):

$$X_k \equiv \text{representative}\{\mathbf{X}_k\}_{k=S,B,O} \equiv rep\{\mathbf{X}_k\}_{k=S,B,O}:$$
$$(X_k \in \mathbf{X}_k) \wedge (\forall X_i, X_j \in \mathbf{X}_k): \quad (3)$$
$$\sum_i r^2(X_k, X_i) = \max\{\sum_i r^2(X_i, X_j) : P[r(X_i,X_j)=0] \leq 0,10\}$$

In the final step, the representative variables of the groups $\mathbf{X}_S$, $\mathbf{X}_B$, and $\mathbf{X}_O$ are placed as independent variables ($X_i$) in a multivariate linear regression model, which has dependent variable the number of commuters ($Y$) and it is described by the following expression (Tsiotas and Polyzos, 2015c):

$$\mathbf{X} \equiv \{X_k, k = 1,...,p\}$$
$$(\mathbf{X} \equiv \mathbf{X}_S \cup \mathbf{X}_B \cup \mathbf{X}_O) \wedge (\mathbf{X}_i \cap \mathbf{X}_j = \emptyset) \wedge (j \neq i, j = \{S, B, O\})$$
$$GCN(V, E):$$
$$Y = f(X_S, X_B, X_O) = b_S \cdot X_S + b_B \cdot X_B + b_O \cdot X_O + c \qquad (4)$$
$$\text{where } X_k \equiv rep\{\mathbf{X}_k\}_{k=S,B,O}:$$
$$X_k \in \mathbf{X}_k \wedge \forall X_i, X_j \in \mathbf{X}_k:$$
$$\sum_i r^2(X_k, X_i) = \max\{\sum_i r^2(X_i, X_j): P[r(X_i, X_j) = 0] \leq 0,10\}$$

## 3. Network topology analysis
### 3.1. Network measures
In the first step of the analysis, we calculate the network measures of the GCN, the results of which are shown in Table 3.

**Table 3**
Network measures of the GCN

| Measure/ Metric | Symbol | Unit | Value |
|---|---|---|---|
| Nodes | $n$ | #[a] | 39 |
| Edges/links | $m$ | # | 71 |
| Self-connections | $n(e_{ii} \in E)$ | # | 0 |
| Isolated nodes | $n_{k=0}$ | # | 0 |
| Components | $\alpha$ | # | 1 |
| Maximum degree | $k_{max}$ | # | 7 |
| Minimum degree | $k_{min}$ | # | 1 |
| Average degree | $\langle k \rangle$ | # | 3.641 |
| Spatial strength | $\langle k_w \rangle$ | km | 322.264 |
| Average nearest neighbor degree | $\langle k_{N(v)} \rangle$ | # | 3.641 |
| Average nearest neighbor strength | $\langle k_{N(v),w} \rangle$ | km | 322.26 |
| Average edge length | $\langle d(e_{ij}) \rangle$ | km | 85.497 |
| Total edge length | $\sum_{ij} d(e_{ij})$ | km | 3334.4 |
| Average path length | $\langle l \rangle$ | # | 4.58 |
| Average weighted path length | $d(\langle l \rangle)$ | km | 389.045 |
| Network diameter (binary) | $d_{bin}(G)$ | # | 14 |
| Network diameter | $d_w(G)$ | km | 1,124.4 |
| Graph density (planar) | $\rho$ | net[d] | 0.640 |
| Graph density (non-planar) | $\rho$ | net | 0.097 |
| Clustering coefficient[c] | $C$ | net | 0.47 |
| Average clustering coefficient[c] | $\langle C \rangle$ | net | 0.422 |
| Modularity | $Q$ | net | 0.566 |

a. number of elements
b. NaN = not a number
c. n/a = not available
d. dimensionless number

By default, the GCN does not have nodes with self-connections ($n(e_{ii} \in E) = 0$), nor isolated nodes ($n_{k=0}$), nor more than one connected component ($\alpha_{GCN}=1$). The maximum network degree is $k_{GCN,max}=7$, whereas the minimum is $k_{GCN,min}=1$ (since GCN is a connected component). Further, the network average degree is $\langle k \rangle_{GCN}=3.641$ and it is close to the mode value in the degree distribution describing the urban road systems, according to the study of Courtat et al. (2010). The average path length generally represents the spatial cost (expressed in number of edges) required to move along a network (Tsiotas and Polyzos, 2015a,b). For the GCN, this cost implies that a path between two random nodes consists of $\langle l \rangle_{GCN}=4.58$ spatial units (edges or steps of separation).

The GCN's average path length ($\langle l \rangle_{GCN}$) is of order of magnitude $\mathcal{O}(\sqrt{n}) = \sqrt{39} \approx 6.245$ that characterizes the average path length $\langle l \rangle_{latt}$ of a lattice with the same number of nodes with the GCN, implying that the GCN may have lattice-like characteristics. In addition, the spatial (kilometric) version of GCN's average path length is $d(\langle l \rangle)_{GCN} = 389.045$km and represents the average kilometric distance required to randomly travel between two network nodes.

Next, the magnitude of the binary network diameter expresses that the most distant binary path that can be traversed in the GCN consists of 14 edges, whereas the most distance kilometric path is $d(GCN)=1,124.40$km (these two diameters are not necessarily the same). The graph density $\rho$ of the GCN, whether it considered as planar graph (i.e. excluding the intersections), equals to $\rho_{1,GCN}=0.64$, whereas for the non-planar case (i.e. including the intersections) equals to $\rho_{2,GCN}=0.097$. Both densities are extremely small compared to the corresponding empirical values of urban road networks (Barthelemy, 2011).

The (global) clustering coefficient of the GCN is $C_{GCN}=0.47$, indicating a good clustering along the network structure. Additionally, the average clustering coefficient (the average of the local clustering coefficients computed for all the network nodes) is equal to $\langle C \rangle_{GCN}=0.422$, which is impressively greater than the corresponding value of a random network ER, namely $\langle C \rangle_{ER} \sim 1/n = 1/39 = 0.026$. This implies that the topology of the GCN is far to be described from a random process.

Finally, the modularity of the GCN is $Q_{GCN}=0.566$, expressing the ability of the network to be divided into communities. This value describes a satisfactory community partitioning capability, which is at least better than the typical road network cases that are of the order $Q_{bipart} < 0.4$.

*3.2. Topological analysis*

The degree of distribution of the GCN is examined at the diagrams ($k$, $n(k)$) shown in Fig.2. These diagrams show a peaked distribution that better fits to a normal ($R^2=0.885$) rather than a power-law ($R^2=0.025$) curve. Also, the peak appearing around the value $k=3$ implies the presence of strong spatial constraints (Barthelemy, 2011) in the structure of the GCN.

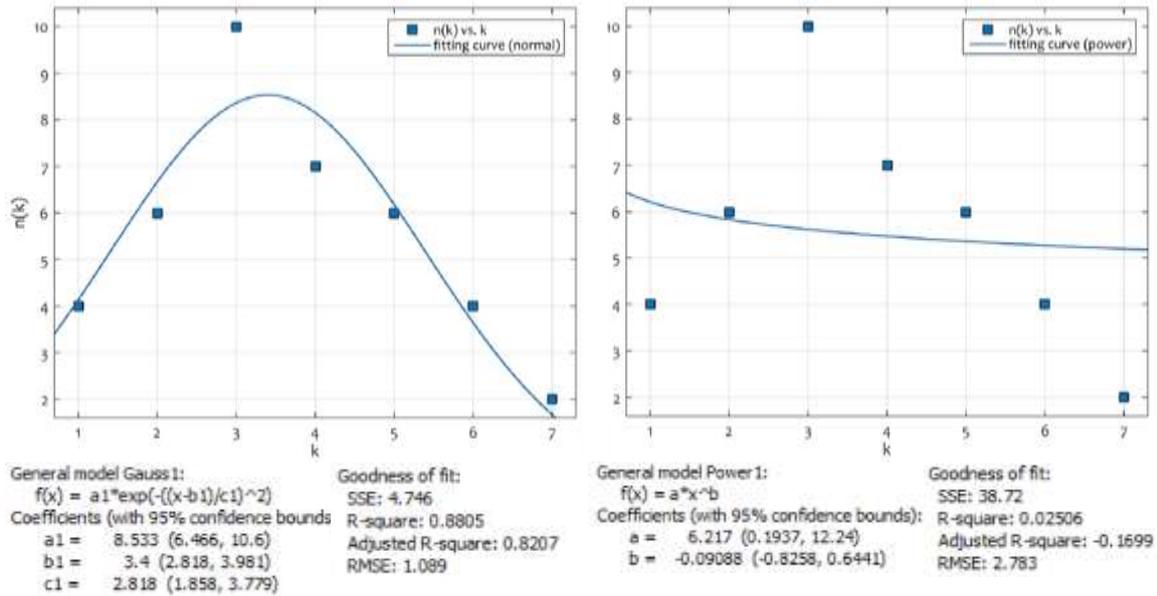

**Figure 2.** Scatter plots ($k$, $n(k)$) of the GCN's degree distribution, with a normal and a power-law fitting curve.

Next the results of the omega ($\omega$) index analysis (Telesford et al., 2011) are shown in Table 4, indicating that the GCN has lattice-like characteristics. This outcome complies with the theory stating that road networks are subjected to intense spatial constraints (Barthelemy, 2011).

**Table 4**
Results of the approximate small-world detection of GCN[*]

| **Measure** | $\langle c \rangle$ | $\langle c \rangle_{latt}$ | $\langle l \rangle$ | $\langle l \rangle_{rand}$ | $\omega^*$ |
|---|---|---|---|---|---|
| GCN | 0.422 | 0.312 | 4.580 | 2.889 | **-0.7218** |
| Indication | | | | | lattice-like characteristics |

[*]. According to relation (1)

At the next step, we compute the basic measures of topology and centrality of the GCN (degree, betweenness, closeness, clustering, modularity, and spatial strength), the results of which are shown in the spatial distributions depicted in the topographic maps of Fig.3. First, the spatial distribution of degree ($k$) (Fig.3a) seems to form a distinctive pattern, where a cluster of hubs (highly connected nodes) is located at the center of the GCN, whereas another (single) hub is located south, at the region of the Peloponnese. The cluster at the central Greece is configured by the prefectures of Larissas (18), Fthiotidas (27), Kozanis (13), Aitoloakarnanias (26), and Ioanninon (15), while the hub in Peloponnese is located at the prefecture of Arkadias (37).

At second, a considerable connectivity is shown by the prefectures of Pella (10) and Thessaloniki (8), in the Northern Greece, as well as by the prefectures of Grevenon (16), Trikalon (19), Karditsas (24), and Artas (21), which are configuring a cluster in the Central Greece. Taking into account that magnitude in degree expresses connectivity and thus the ability of nodes to communicate in the network, it seems that the spatial distribution of degree (Fig.3a) points out those nodes in the GCN that have a connectivity advantage over the others. This advantage is due to the geographical arrangement of the Greek prefectures, which favors the development of links in central places rather than in regional areas.

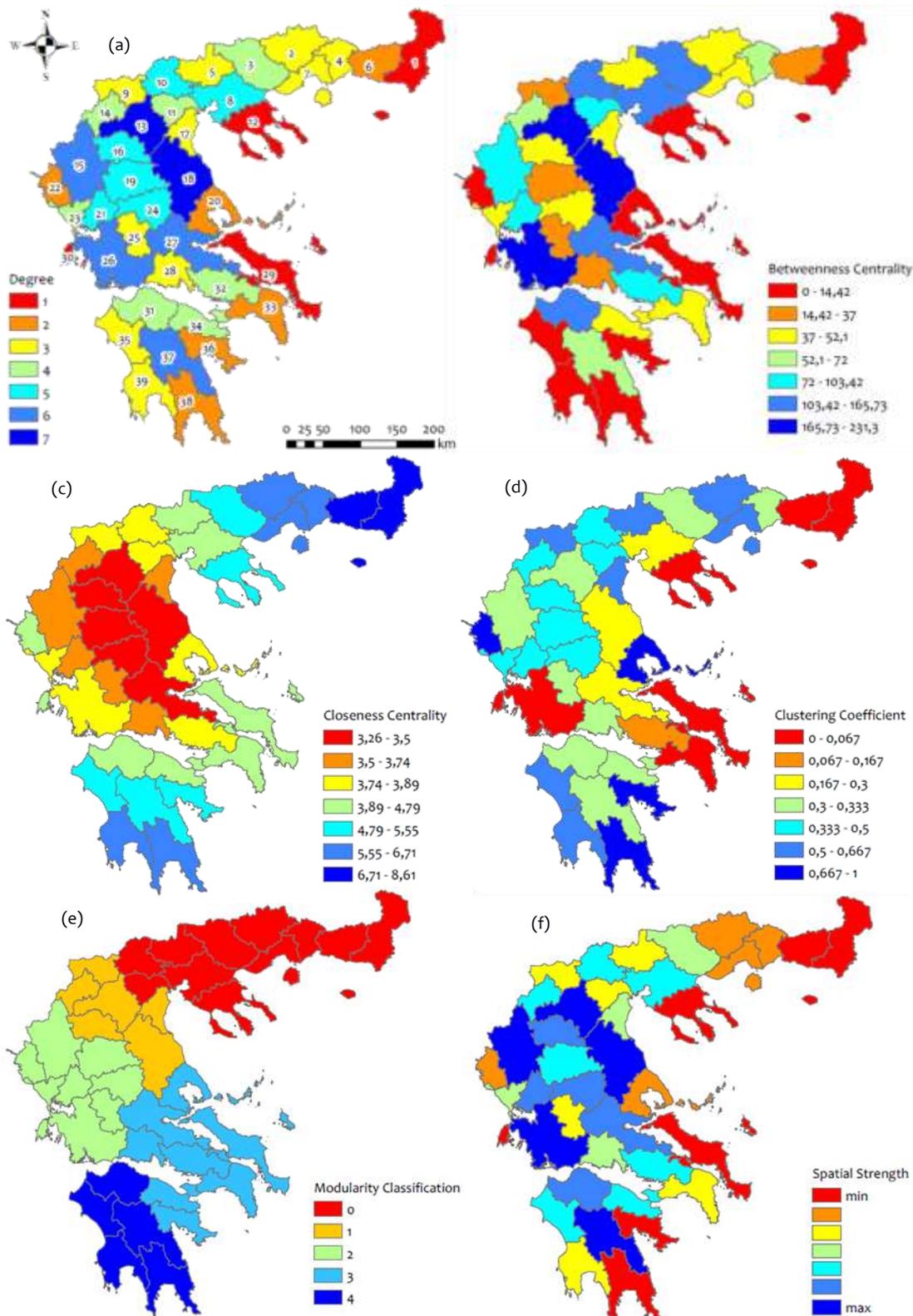

**Figure 3.** Spatial distribution of the GCN's node measures: (a) Degree (b) Betweenness (c) Closeness (d) Clustering (e) Modularity classification, and (f) Spatial strength.

Next, the spatial distribution of the betweenness centrality (Fig.3b) shows a higher intensity at the eastern part of the country, where more upgraded road infrastructure occurs (Tsiotas et al., 2012). On the other hand, the closeness centrality spatial distribution (Fig.3c) shows small values in borderline regions (Eastern Macedonia, Thrace, West

Peloponnese), whereas its higher values are also central, verifying the clear advantage that central areas have in spatial networks. In terms of clustering coefficient, the spatial distribution of this measure (Fig.3d) shows higher values to be arranged mainly at the periphery, namely at the prefectures of Heleias (35), Messhnias (39), Lakonias (38) and Argolidas (36), in the Peloponnese, at the prefectures of Thesprotias (22) and Magnessias (20) in the central country, and at the prefectures of Pierias (17), Florinas (9), Kilkis (5), Dramas (2), and Kavalas (7) in the northern Greece. This configuration generally implies that the peripheral regions are more likely to be connected with interconnected neighbors, describing a privilege to receive more coherent information by their network. However, this privilege may also suggest a disadvantage because it implies the dependence of these regions on their neighbors, in terms of interaction. For the GCN, the ability to access high clustering network nodes depends on their neighbors' road infrastructure quality, which may have similar qualitative characteristics, due to the high degree of their interdependence. At next, the spatial distribution of the modularity classification (Fig.3e) appears to be consistent with the theory (Guimera et al., 2005; Kaluza et al., 2010; Barthelemy, 2011). In particular, this distribution illustrates a distinct partition into geographical areas, which is expected for a lattice network, such as the interregional GCN. Finally, the distribution of the spatial strength (Fig.3f) appears to be more intensive in center places, forming a cluster consisting of the prefectures of Fthiotidas (27), Larissas (18), Kozanis (13), Ioanninon (15), Artas (21), Aitoloakarnanias (26), and Arkadias (37).

In terms of time distances, we examine the differences between two time snapshots of the years 1988 and 2010, for the measures of degree and closeness centrality (Fig.4a,b). This diachronic change may provide insights about the connectivity of the Greek interregional road network and about potential improvements in the road infrastructures occurred in the meanwhile. The comparison of these time snapshots shows an almost identical network structure, except from the prefectures of Aitoloakarnanias (26) and Achaias (31), which increased their degree status. This change is obviously due to the construction of the Rio-Antirion bridge, in the year 2004, which connected the prefectures of Achaias and Aitoloakarnanias and provided direct access to them (Tsiotas et al., 2012).

At next, the spatial distribution of closeness centrality for the years 1988 vs. 2010 is shown in Fig.4b. First, the 1988 map describes the accessibility of the road transport network in that period, which appears generally inaccessible for the borderline prefectures. For the year 1988, the most privileged in connectivity were the prefectures of Kozanis (13), Larissas (18), Ioanninon (15), and of Arkadias (37), obviously due to their central geographical location, while in a secondary position were the prefectures of Central Greece and of Thessaloniki. In particular, the prefecture of Thessaloniki seems to have played a remarkable role in the network's accessibility at that time, since it was a metropolitan core of the country since 1988 (Tsiotas et al., 2012). The comparison between the 1988 and the 2010 maps of the GCN's closeness centrality outlines the qualitative changes occurred in the road network infrastructures of Greece in that period (Tsiotas et al., 2012). Obviously, there was an upgrade in the structure of this network that period (1988-2010), which favored the broader region of Central Greece that included prefectures with highest accessibility. A remarkable reduction appears in the accessibility classification of the prefectures of Rodopis (6) and Voiotias (32), each of which is attributed to different reasons. On the one hand, the decrease in the case of Rodopi is apparently due to its geographical location, since the accessibility of its road network has increased in absolute terms. On the other hand, the decrease in the case of Voiotias is probably related to the construction of the Rio-Antirion bridge, which practically "disconnected" Voiotia with the prefectures of the Peloponnese region (31, 35, 39, 38, 37, 36, 34) for those paths leading to the Central Greece (Tsiotas et al., 2012).

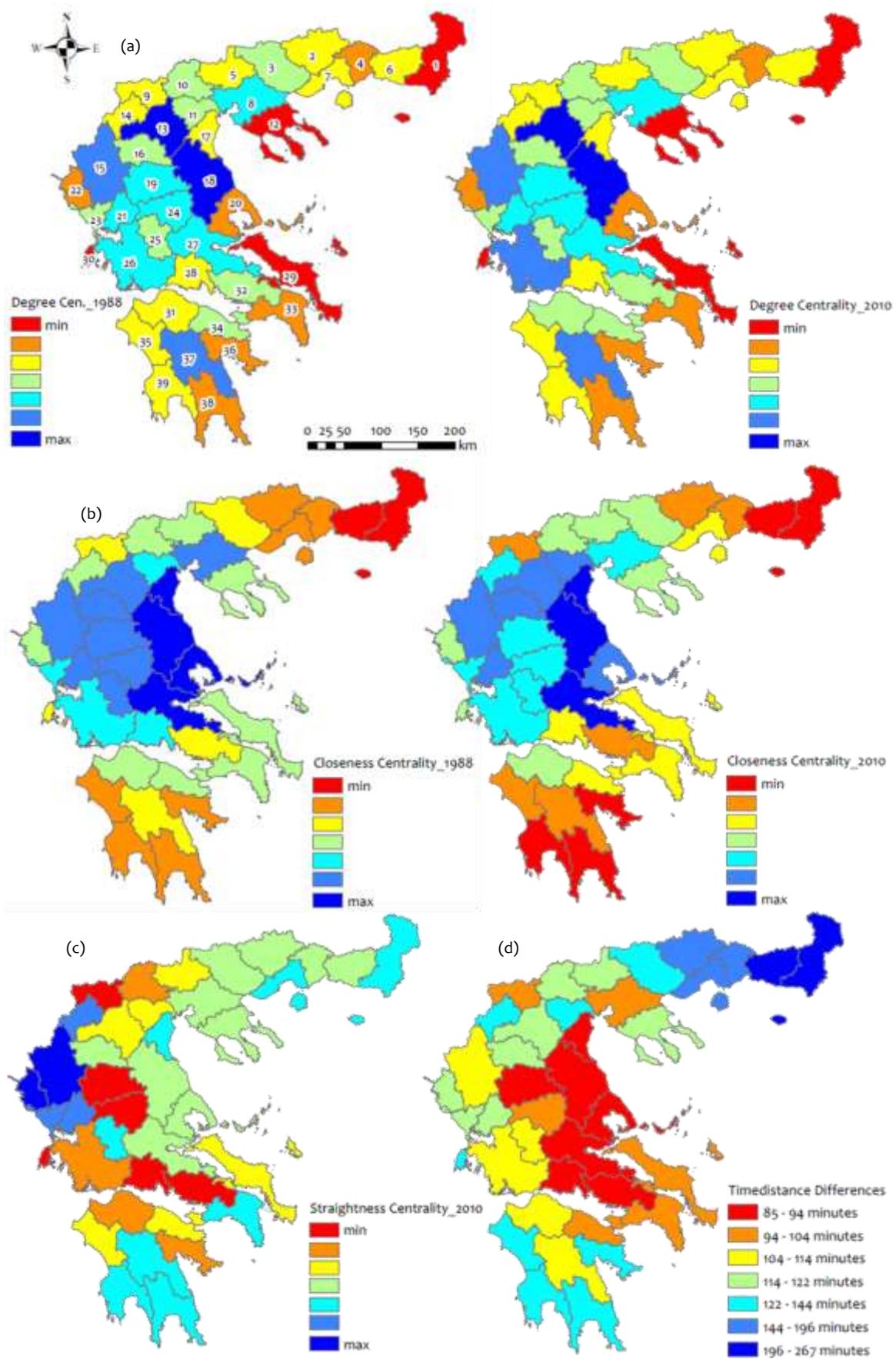

**Figure 4.** Spatial distribution of the GCN's (a) degree and (b) closeness centrality, for the years 1988 vs. 2010 (color gradation depicts classification and not absolute centrality values), (c) spatial distribution of the GCN's straightness centrality in 2010, and (d) changes in time distances for the years 1988 vs. 2010 (source: Tsiotas, 2012).

Furthermore, the spatial distribution of the straightness centrality (Fig.4c) expresses the deviation from direct access (of a route) and thus the effectiveness of the road network. Thus, this measure may also be used as an indicator of the quality of the GCN's road infrastructures (Tsiotas et al., 2012). Therefore, color gradations in the straightness centrality map illustrate the benefit that regions received from the road infrastructures upgrade during the period 1988-2010. Within this context, the prefectures benefited the most from the road infrastructure projects in Greece were Ioanninon (15), Thesprotias (22), Artas (21), Prevezas (23), and Kastorias (14). Their geographic position indicates that they have benefited from both the Rio-Antirio Bridge construction and from the Egnatia Motorway construction that was conducted during the period 1994-2009 and improved the connectivity of the intermediate prefectures between Thesprotias (22) and Evrou (1). At a second level, higher straightness centrality scores are observed for:
- The prefectures-cluster of Kastorias (14), Prevezas (23), and Evrytanias (25), which are adjacent to the previously mentioned most benefited prefectures.
- The prefectures-cluster of Arcadias (37), Laconias (38), and Messhnias (39), in the region of Peloponnese, who apparently benefited more from the Rio-Antirio link,
- The prefecture of Attikis (33), which was obviously favored by the total road upgrading projects, as well as
- The prefectures of Kavalas (7) and Evrou (1), in northern Greece, which appeared to have been favored by the construction of the Egnatia Motorway.

At next, the geographical distribution of the time-distance differences (Fig.6d) illustrates the prefectures that benefited the most from the Greek transport policy in terms of the road infrastructure works conducted at the period 1988-2010 (Tsiotas, 2012). According to the map, the time-distance distribution shows a distinct spatial grouping with highest values in the periphery and the smallest in the center. In particular, the prefectures that benefited the most are the border of Evrou (1) and Rodopis (6) and, secondly, the perfectures of Xanthis (4), Kavalas (7), and Dramas (2). The prefectures following this level, are Serron (3), Hemathias (11), and Florinas (9), in Northern Greece, the prefectures of Prevezas (23) and Lefkados (30), in the West Greece, and the prefectures of Heleias (35), Messinias (39), Lakonias (38), and Argolidas (36) in the Peloponnese.

The last part of the GCN's network analysis shows the correlations between the betweenness centrality $C^b(k)$ and the spatial strength $s(k)$ with respect to the network degree $k$ (Fig.7). According to the fitting curves $(k, \langle C^b|_{k=k_i} \rangle)$ and $(k, \langle s|_{k=k_i} \rangle)$, a remarkable linearity for both cases exists, with coefficients of determination $R^2_{C^b,k}=0.96$ and $R^2_{s,k}=0.906$, respectively.

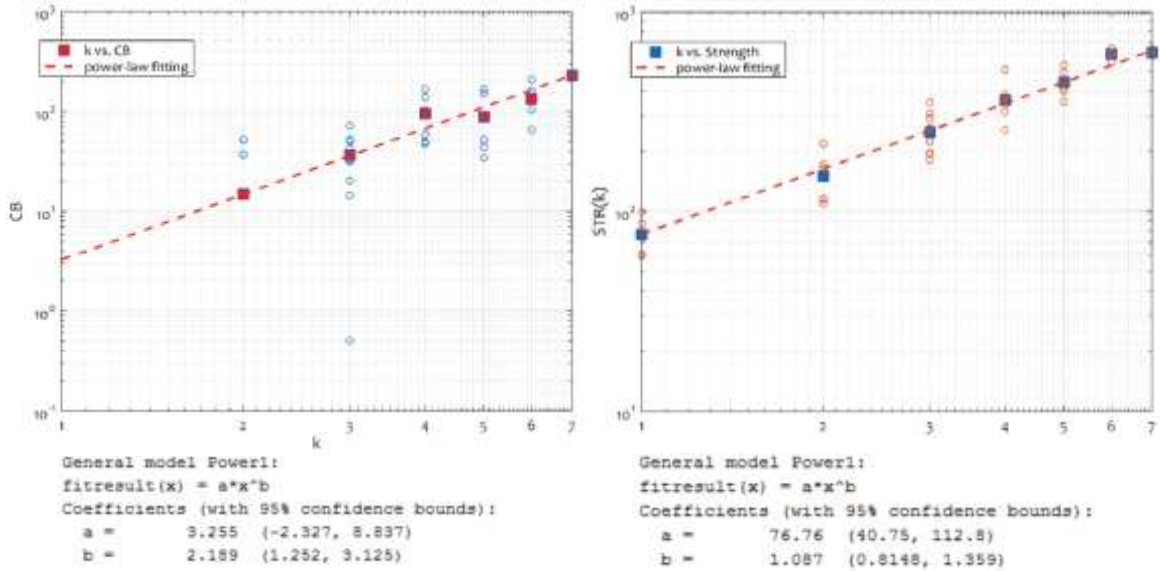

**Figure 5.** Scatter Plots showing the correlations between (left) the betweenness centrality and network degree (Cb, k) and (right) the spatial power and network degree (s, k), for the GCN (square marks represent the average values per degree class).

The relationship $\langle C^b|_{k=k_i}\rangle = f(k)$, between the degree ($k$) and the mean value of the betweenness centrality per degree $\langle C^b|_{k=k_i}\rangle$, where $i=2,3,...,7$, has a power-law exponent $\beta_{GCN}=1.94$ and expresses that the strong connecting nodes in the network (hubs) undertake the largest load of its traffic. On the other hand, the exponent $\beta_{GCN}=1.156$ of the relationship $\langle s|_{k=k_i}\rangle = f(k)$, between the degree $k$ and average spatial strength $\langle s|_{k=k_i}\rangle$, where $i=2,3,...,7$, is close to the unit (~ 1) and indicates an almost linear relation between the variables, which implies a homogeneity regarding the undertake of distant traffic in the GCN.

Finally, Fig.6 shows the correlation ($k$,$C(k)$) between the variables $k$ (node degree) and C (clustering coefficient) of the GCN.

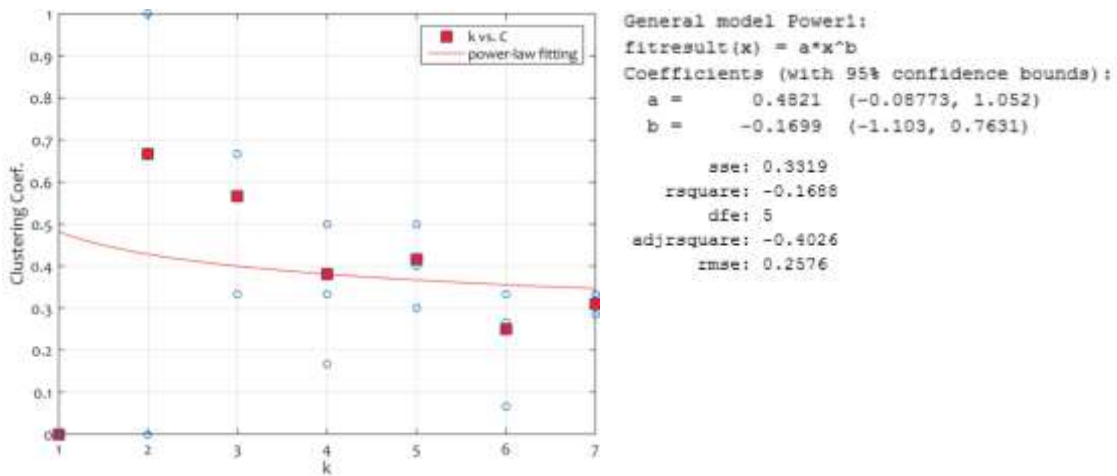

**Figure 6.** Correlation between the clustering coefficient and network degree (C, k) of the GCN. The shape of the scatter plot indicates a logarithmic decrease as the $k$ values increase.

The relationship $C=f(k)$ (Fig.6) indicates a logarithmic decrease of the GCN's clustering by increasing the degree ($k$) values, which is consistent with the common research practice (Sen et al., 2003; Barthelemy, 2011). This relationship describes that as the connectivity of a node increases in the network, the possibility of this node to be associated with interconnected neighbors is reduced.

## 4. Empirical Analysis

Table 5 shows the results of the groups' ($X_S$, $X_B$, and $X_O$) representative variables selection, according to the algorithm shown in relation (3). In this table, the ranking (hierarchy) of the variables calculated within the groups is shown in the first column (Position within group), whereas the ranking calculated for the total (Global calculations) of variables are shown in the columns with the indication "Rank".

**Table 5**
Results of the analysis of choosing the groups representatives

| Position within group | Structural variables – $X_S$ | | | | Functional variables – $X_B$ | | | | Ontological cariables – $X_O$ | | | |
|---|---|---|---|---|---|---|---|---|---|---|---|---|
| | Within-groups | | Global | | Within-groups | | Global | | Within-groups | | Global | |
| | Variable | $\Sigma_S(r^2)$ [a] | Rank[b] | $\Sigma(r^2)$ | Variable | $\Sigma_B(r^2)$ | Rank | $\Sigma(r^2)$ | Variable | $\Sigma_O(r^2)$ | Rank | $\Sigma(r^2)$ |
| 1 | **$S_6$** | **2.746** | **2** | **12.116** | $Y$[c] | 5.285 | 1 | 12.181 | $O_2$ | **4.091** | 9 | 10.451 |
| 2 | $S_5$ | 2.738 | 3 | 12.058 | **$B_6$** | **5.190** | **4** | **11.905** | $O_6$ | 3.990 | 7 | 11.703 |
| 3 | $S_{10}$ | 2.493 | 11 | 7.877 | $B_7$ | 5.152 | 6 | 11.730 | $O_7$ | 3.983 | **5** | **11.805** |
| 4 | $S_1$ | 2.448 | 23 | 2.596 | $B_8$ | 5.048 | 8 | 11.428 | $O_4$ | 3.256 | 13 | 6.443 |
| 5 | $S_2$ | 2.188 | 22 | 2.945 | $B_3$ | 3.910 | 10 | 9.400 | $O_9$ | 2.612 | 19 | 3.859 |
| 6 | $S_9$ | 1.777 | 25 | 2.137 | $B_5$ | 3.362 | 12 | 7.177 | $O_1$ | 2.507 | 14 | 5.169 |
| 7 | $S_4$ | 1.753 | 27 | 1.838 | $B_4$ | 2.622 | 16 | 4.883 | $O_3$ | 2.236 | 21 | 3.307 |
| 8 | $S_{11}$ | 1.576 | 17 | 4.404 | $B_1$ | 1.628 | 20 | 3.467 | $O_5$ | 1.935 | 18 | 3.898 |
| 9 | $S_8$ | 1.570 | 28 | 1.790 | $B_2$ | 1.000 | 30 | 1.206 | $O_8$ | 1.849 | 15 | 5.080 |
| 10 | $S_3$ | 1.569 | 24 | 2.169 | | | | | $O_{10}$ | 1.368 | 29 | 1.458 |
| 11 | $S_7$ | 1.158 | 26 | 1.931 | | | | | | | | |

a. Sum of squares of correlation coefficients
b. Relative position of the variable in the total ranking (global calculations)
c. This variable is exempted because is considered as response variable in the model

According to the results of Table 5, the representative variables resulting from the within-group calculations are the structural variable $S_6$ (population), the functional variable $B_6$ (number of vehicles), and the ontological variable $O_2$ (educational index). The corresponding results for the global analysis (in the total of variables) are slightly different, resulting to the variables $S_6$, $B_6$, and $O_7$ (number of car accidents) instead of the $O_2$, respectively. Taking into account that the position of the $O_2$ variable for the within-class analysis is three steps lower than its position in the global analysis, for the sake of completeness we use in the analysis three (instead of one) representative groups of variables, namely ($S_6$, $B_6$, $O_2$), ($S_6$, $B_6$, $O_7$), and ($S_6$, $B_6$, $O_6$), which are produced by the global ranking $O_7$, $O_6$, and $O_2$.

An interesting observation resulting from Table 5, is that the dependent variable $Y$ (number of commuters) is placed for both the within-class and global analysis in the first ranking, having the highest correlation coefficients sum of squares. This result is rationale because the independent variables have been chosen on the basis of their theoretical relevance to the commuting phenomenon and thus their direct or indirect correlation with the response variable is somewhere expected. Another interesting observation is that the population variable ($S_6$) shows the greatest correlation with the other independent variables, illustrating a gravitational pattern of the commuting phenomenon (Tsiotas and Polyzos, 2015b).

In the last step, the variables emerged as representatives of the groups $\mathbf{X}_S$, $\mathbf{X}_B$, and $\mathbf{X}_O$ are entered as independent variables ($X_i$) into a multivariate linear regression model, with dependent variable the number of commuters ($Y$). Table 6 shows the results of this analysis, which is applied to the three different sets of representative variables and it produces three distinct linear regression models, namely $Y_1=f(S_6,B_6,O_2)$, $Y_2=f(S_6,B_6,O_7)$, and $Y_3=f(S_6,B_6,O_6)$.

**Table 6**
Results of the multivariate linear regression analysis

| Model[a] | | | Non-standardized coefficients | | Standardized coefficients | | Sig.[e] |
|---|---|---|---|---|---|---|---|
| Model info | | Predictor variables | $b$[b] | St. Error (S.E.) | $b$[c] | $t$[d] | |
| ($Y_1$) Model: (constant), $S_6$, $B_6$, $O_2$ | | | | | | | |
| $R$[f] | 0.999 | (constant) | -881.91 | 174.10 | | -5.066 | 0.000 |
| $R^2$[g] | 0.998 | $S_6$ | 0.011 | 0.002 | 0.600 | 4.840 | 0.000 |
| S.E of the estimation | 600.75 | $B_6$ | 0.011 | 0.004 | 0.345 | 3.003 | 0.005 |
| | | $O_2$ | 41.422 | 11.729 | 0.065 | 3.532 | 0.001 |
| ($Y_2$) Model: (constant), $S_6$, $B_6$, $O_7$ | | | | | | | |
| $R$ | 0.998 | (constant) | -674.49 | 212.47 | | -3.174 | 0.003 |
| $R^2$ | 0.997 | $S_6$ | 0.015 | 0.002 | 0.816 | 6.151 | 0.000 |
| S.E of the estimation | 693.20 | $B_6$ | 0.02 | 0.005 | 0.049 | 0.335 | 0.740 |
| | | $O_7$ | 1.198 | 1.482 | 0.134 | 0.809 | 0.424 |
| ($Y_3$) Model: (constant), $S_6$, $B_6$, $O_6$ | | | | | | | |
| $R$ | 0.999 | (constant) | -532.81 | 199.46 | | -2.671 | 0.011 |
| $R^2$ | 0.997 | $S_6$ | 0.016 | 0.002 | 0.834 | 7.991 | 0.000 |
| S.E of the estimation | 643.41 | $B_6$ | 0.021 | 0.007 | 0.627 | 2.796 | 0.008 |
| | | $O_6$ | -6.337 | 2.508 | -0.462 | -2.527 | 0.016 |

a. Enter method (including all variables entered in the model)
b. Non-standardized coefficients *beta* of the model
c. Standardized coefficients *beta* of the model
d. *t*-statistic for the coefficients' significance testing
e. 2-tailed significance
f. Multiple correlation coefficient
g. Coefficient of determination

The determination coefficients values ($R_2$), in Table 6, express that the three models $Y_1$, $Y_2$, and $Y_3$ have almost excellent ability to describe the variability of the response variable (number of commuters). This determination ability is also deduced from the sums of the standardized regression coefficients, which are close to the unit for each case, implying the absence of significant collinearity between the variables (Tsiotas and Polyzos, 2015a). This observation signifies the utility of the proposed multivariate linear regression approach, especially when taking into consideration that it is based on the ~ 10% of the available information (i.e. on 3 of the 29 available variables). The selection of the representative variables assigns a systolic property to the methodological approach applied in this paper, which complies with the conceptual framework of the term "network" as described by Tsiotas and Polyzos (2015c).

Overall, the results of Table 6 indicate that the population variable ($S_6$) is the most important determinant in the commuting phenomenon in Greece. This observation is consistent with the theory (Polyzos, 2011; Polyzos et al., 2014, 2015), highlighting the gravitational dimension of commuting, since the contribution rate of the $S_6$ variable (as it is induced from standardized beta coefficients in the models) ranges between 60-83%. Further, the presence of the $B_6$ variable (number of private cars) in the models implies that the use of private cars by workers plays a crucial role in the interregional commuting. In the interregional scale, the use of alternative modes of transport (bus, train) does not appear to be a determinative factor for commuting, obviously due to the fact that in the interregional level time-distances are already stretched (the maximum possible) and thus there is no option to add more delays by using public transport.

In the $Y_1$ model, the contribution of variable $B_6$ (number of vehicles) is about 34.5%, whether the contribution of the variable $O_2$ (educational level) is limited to 6.5%. In the $Y_2$ model, both the contribution of the variable $B_6$ and $O_7$ (number of car accidents) appears statistically insignificant, implying that these two variables are not related in common with the commuting phenomenon.

Finally, in the $Y_3$ model, the contribution of variable $B_6$ is about 62.7%, but together with the population variable ($S_6$) appear to compete with the $O_6$ variable (specialization in transportation – transportations GDP), as it is extracted by their positive and negative signs. This model also highlights a gravitational pattern of commuting and it additionally implies that specialized in transportation prefectures tend to maintain a large amount of commuting activity within their urban boundaries and thus to limit interregional transportation and its produced product.

## 5. Conclusions

This article studied the topology of the interregional commuting network in Greece (GCN) by using the network paradigm. The GCN, consisting of 39 non-insular prefectures, was modeled into a complex network and it was studied using measures and methods of complex network analysis and empirical techniques. The purpose of the study was to detect the structural characteristics of the GCN and to interpret how this network is related to the regional development. Towards this direction, the effect of the spatial constraints in the structure GCN was evident, as it was supported by the following observations:

- The degree distributions were peaked deviating from the power-law pattern that describes networks with smooth spatial constraints.
- The estimation of the omega (ω) index, which is used to approximately detect the small-world property and the lattice-like and random-like topology in networks, showed the existence of lattice-like characteristics in the GCN.
- Central in geography places in the GCN showed highest values in terms of degree and betweenness centrality.
- The GCN's communities produced by the modularity optimization showed geographical consistency και
- Correlations between degree and betweenness centrality ($k,C^b$) and degree and spatial strength ($k,s$) followed *power-law* patterns, whereas this with clustering ($k, C$) showed a logarithmic decay.
- Large variability was detected in betweenness centrality ($C^b$), illustrating that this measure has a clear geographic configuration, tending to be identified with the gravitational center of the network.

Despite the detection of spatial constraints on the GCN topology, the form of the relation $s=f(k)$ showed the existence of long-distance connections that are detected when the power-law exponent is greater than one ($β >1$). This is obviously due to the construction rule of the GCN, whose edges have conceptual (expressing potential of road connectivity between the Greek prefectures) and not physical interpretation.

Further, the comparison of the centrality measures calculated for two different timeframes (1988 vs. 2010) gave insights about the influence that some major road infrastructure projects have on the GCN. Detected changes uncovered the prefectures that benefited the most from the transport infrastructure policy during the period 1988-2010. Overall, the road transport infrastructure policy of that period in Greece appeared to follow a planning favoring the borderline and peripheral regions, targeting to eliminate geographical inequalities and to promote regional development.

In the part of the empirical analysis, a multivariate linear regression model was constructed, based on a variables' classification inspired from a conceptual model describing the term "network" in Network Science. The analysis was conducted on 30 vector variables including regional values for each network attribute, which were grouped by their thematic relevance. Three groups of three representative variables were selected from each category, according to their within-group and global sums of correlations, which afterwards were used as independent variables to construct the regression models ($Y_1$, $Y_2$ and $Y_3$). The results of this analysis highlighted the gravitational pattern of commuting, since the population variable $S_6$ had a contribution, in all models, about 60-83%. The presence of the other (functional and ontological) variables in the models showed the importance that the private transport plays in the commuting (ranging from 5-63% in the three models), whereas the choice of alternative transport modes (bus, train) appeared insignificant. Also, the analysis showed that, in the interregional scale, the commuting has also an ontological aspect where the educational level was the most important component (factor) affecting the phenomenon at a 6.5% level. Finally, the analysis highlighted that specialized in transportation prefectures tend to maintain a large amount of commuting activity within their urban boundaries and thus to limit interregional transportation and its produced product.

Overall, this paper highlighted the effectiveness of complex network analysis in the modeling of systems of regional economy, such as the systems of spatial interaction and the transportation networks, and it supported the use of the network paradigm in regional research.